\documentclass[12pt,preprint]{emulateapj}
  
\usepackage{apjfonts}

\shorttitle{AGN Fraction at z $\sim$ 0.3}
\shortauthors{Finkelstein et al.}

\newcommand{\sol}{$_{\odot}$}
\newcommand{\lya}{Ly$\alpha$}

\def\arcs{\hbox{$^{\prime\prime}$}}

\begin{document}
\title{A Plethora of AGN Among \lya~Galaxies at Low Redshift\altaffilmark{1}}

\author{Steven   L.    Finkelstein\altaffilmark{2}, Seth H. Cohen\altaffilmark{3}, Sangeeta Malhotra\altaffilmark{3}, James E. Rhoads\altaffilmark{3}, Casey Papovich\altaffilmark{2}, ZhenYa Zheng\altaffilmark{4} \& Jun-Xian Wang\altaffilmark{4}}   
\altaffiltext{1}{Observations reported here were obtained at the MMT Observatory, a joint facility of the University of Arizona and the Smithsonian Institution.  These observations were also based in part on observations made with the NASA Galaxy Evolution Explorer.  GALEX is operated for NASA by the California Institute of Technology under NASA contract NAS5-98034.}
\altaffiltext{2}{George P. and Cynthia W. Mitchell Institute for Fundamental Physics and Astronomy, Department of Physics, Texas A\&M University, College Station, TX 77843; stevenf@physics.tamu.edu}
\altaffiltext{3}{School of Earth and Space Exploration,  Arizona  State University,  Tempe, AZ  85287} 
\altaffiltext{4}{Center for Astrophysics, University of Science and Technology of China, Hefei, Anhui 230026, P. R. China}

\begin{abstract}
We investigate the fraction of z $\sim$ 0.3 Lyman alpha emitting galaxies (LAEs) which host active galactic nucleus activity, which is typically from 1 -- 5\% at 2 $\lesssim$ z $\lesssim$ 6.  Using optical spectroscopy of 23 LAEs at 0.2 $\leq$ z $\leq$ 0.45 selected with GALEX UV data, we probe for AGN with a variety of methods, including line widths, diagnostic line ratios, high-ionization emission, infrared activity and X-ray luminosity.  We found that our sample of low-redshift LAEs has an AGN fraction of 43 $^{+18}_{-26}$\%, significantly higher than at high redshift.  While previous results have shown that low-redshift LAEs have a lower space density than their high-redshift analogs, these results show that star-forming LAEs at low-redshift are rarer still.  Accounting for differences in available AGN classification methods, we conclude that rest-frame optical spectroscopy is necessary to identify low-luminosity AGNs in LAEs at high redshift, and that limits on the X-ray flux are necessary to determine the extent to which the AGN contaminates the \lya~fluxes.
\end{abstract}


\section{Introduction}
Lyman alpha (Ly$\alpha$) emitting galaxies (LAEs) have been used for years as probes of the high-redshift universe \citep[e.g.,][]{cowie98,rhoads00,malhotra02,ouchi05,ouchi08,gawiser06b,gronwall07,pirzkal07,nilsson07a,nilsson09,finkelstein07,finkelstein08,finkelstein09a}.  These studies have shown that LAEs are fairly common from 2 $\leq$ z $\leq$ 7, and more interestingly, their properties do not evolve strongly with time, with stellar populations and observed Ly$\alpha$ luminosity functions which are roughly constant over this redshift range.  \citet{deharveng08} discovered $\sim$ 100 low-redshift LAEs from 0.2 $\leq$ z $\leq$ 0.45 with the Galaxy Evolution Explorer (GALEX).  They ruled out objects with \lya~line widths $>$ 1200 km s$^{-1}$ as broad-lined active galactic nuclei (AGN), although they typically could not rule out narrow-lined AGN, as the C\,{\sc iii}] or C\,{\sc iv} diagnostic lines were either too faint, or located in a noisy part of the spectrum.  They found that the Ly$\alpha$ luminosity function of these objects was substantially different than at high-redshift, with low-redshift LAEs having a lower surface density and being typically less luminous in the \lya~line.  

In \citet{finkelstein09c}, we studied the physical properties of 30 of these LAEs which had optical counterparts in the Extended Chandra Deep Field -- South (CDF--S) and Extended Groth Strip (EGS), and found them to be significantly older and more massive than their high-redshift counterparts, with typical ages of a few Gyr, and stellar masses of $\sim$ 10$^{10}$ M\sol.  A few of these objects were poorly fit, which could imply AGN contamination, but with only one object detected in X-rays, we could not discern AGNs from star-forming galaxies.  In this Letter, we present an optical spectroscopic study of 23 of these low-redshift LAEs in order to determine the AGN fraction in LAEs at low redshift.  While Deharveng et al.\ have already ruled out most broad-lined AGN, optical spectroscopy will enable us to search for narrow-lined AGN.  Where applicable, we assume H$_{o}$ = 70 km s$^{-1}$ Mpc$^{-1}$, $\Omega_{m}$ = 0.3 and $\Omega_{\Lambda}$ = 0.7 \citep[c.f.][]{spergel07}.

\section{Data}
\subsection{Observations and Data Reduction}
We observed 23 low-redshift LAEs in the EGS using Hectospec at the 6.5m MMT, which has a 1$^{\circ}$ diameter field-of-view \citep{fabricant05}.  We obtained 2 hours of spectroscopy on 19 March 2009, with spectral coverage from $\sim$ 3650 - 9200 \AA, and spectral resolution of $\sim$ 5 \AA.  We used the External SPECROAD \footnote[5]{http://iparrizar.mnstate.edu/$\sim$juan/research/ESPECROAD/index.php} pipeline for data reduction (developed by Juan Cabanela; private communication).  This pipeline performs standard spectroscopic reductions, including bias, dark and flat correction, and wavelength calibration (rms residuals $\approx$ 0.1 \AA).  We flux-calibrated the spectra using observations of Sloan Digital Sky Survey (SDSS) F-type stars.  Further details on the observations and data reduction, as well as tabulated line fluxes, will be presented in a future paper (Finkelstein et al.\ 2009, in prep).

\subsection{Line Fitting}
The 23 LAEs varied in redshift from 0.2 $\lesssim$ z $\lesssim$ 0.45, thus we found 31 possible spectral lines which we could detect, from Mg\,{\sc ii} $\lambda\lambda$ 2796, 2803 to [S\,{\sc ii}] $\lambda$6733.  We examined the expected position (using the redshift from \citet{deharveng08} as a first estimate) of each of the expected lines in each spectrum, and noted whether it was detected.  We fit a Gaussian curve to each detected line, using the MPFIT IDL software package\footnote[6]{http://cow.physics.wisc.edu/$\sim$craigm/idl/idl.html}, obtaining the continuum flux, central wavelength, Gaussian $\sigma$ and line flux.  Errors on each of these parameters were estimated via Monte Carlo simulations by varying each spectral data point within its photometric uncertainty.  Table 1 lists the object redshifts as determined from this spectroscopy, which are consistent with (and refine) those published in \citet{deharveng08}.
\begin{figure}
\epsscale{1.0}
\plotone{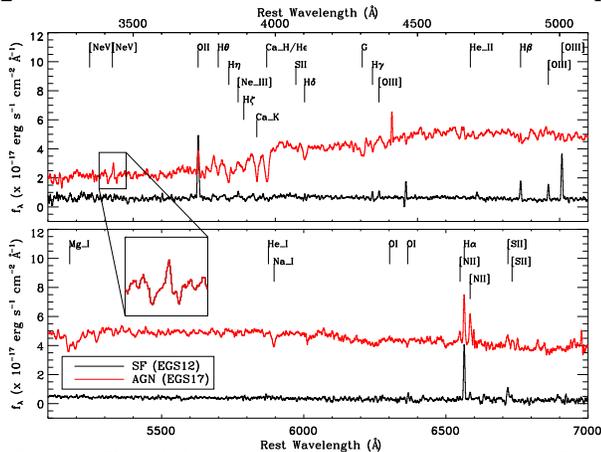}
\vspace{0.2cm}
\caption{The full Hectospec spectra of two of our objects.  The black spectrum denotes a galaxy which we determined to be dominated by star formation, while the red spectrum denotes a galaxy which is likely dominated by AGN activity, as evidenced by the [Ne\,{\sc v}] emission and diagnostic line ratios.}\label{fig:fullspec}
\end{figure}

\section{Results}
Figure~\ref{fig:fullspec} shows spectra of two of the observed LAEs, one which we classify as an AGN, and one which is dominated by star formation.  While some features are obviously different in the AGN, many still appear the same, thus one requires numerous diagnostics to investigate the presence of an AGN.  Combining our spectra with the public data from the All--Wavelength Extended Groth Strip International Survey (AEGIS; \citet{davis07}), we perform five tests to search for AGN.  We first search for broad-lined AGN by examining the widths of permitted lines.  We then search our dataset for emission lines from ions which require high energies to ionize their current state, such as [Ne\,{\sc v}] (126.2 eV).  Next, we use ratios of emission lines to classify them using diagnostic plots from \citet*[hereafter BPT]{baldwin81}.  We then use infrared data from the {\it Spitzer Space Telescope} to probe for warm dust re-emission from absorbed high-energy photons.  Finally, we use archival {\it Chandra X-ray Observatory} data to search for X-ray emission, which could be indicative of an accretion disk around a central black hole. In the following section, the object IDs refer to those defined in \citet{finkelstein09c}.  In Table 1 we list these IDs along with the IDs from \citet{deharveng08}.

\subsection{Line Widths}
Lines from the broad-line region of AGNs have velocities $\gtrsim$ 1000 km s$^{-1}$, while narrow-line region lines have widths $\lesssim$ 500 km s$^{-1}$.  Depending on the orientation, one may only see one of these regions, leading to two separate classification of AGNs: Broad-lined AGN (BLAGN, also called Type 1 or Seyfert 1), and narrow-lined AGN (NLAGN; Type 2; Seyfert 2).  For each object, we measured the characteristic line velocity widths for all detectable, permitted transitions of H\,{\sc i}, He\,{\sc i} and He\,{\sc ii}.  Only EGS24 has a velocity width $>$ 1000 km s$^{-1}$, with v = 1064 $\pm$ 17 km s$^{-1}$, and thus likely contains a BLAGN.  The remaining objects have v $<$ 300 km s$^{-1}$, implying that if an AGN is present in the rest of the sample, it is narrow-lined.  

\subsection{High-Ionization Emission}
[Ne\,{\sc v}] and He\,{\sc ii} are two of the strongest indicators of AGN activity to which our data are sensitive.  We detect [Ne\,{\sc v}] $\lambda$3347 or [Ne\,{\sc v}] $\lambda$3427 in EGS16, EGS17, EGS18 and EGS24 at $>$ 3 $\sigma$ significance, and in EGS14 and EGS20, at 2.9 and 2.0 $\sigma$, respectively.  The He\,{\sc ii} $\lambda$4686 line is also an AGN indicator, and this line is observed at $\geq$ 3 $\sigma$ significance in EGS2, and at 2.6 $\sigma$ in EGS24.  We note that although He\,{\sc ii} emission can also come from the winds of Wolf-Rayet stars, these lines are typically broad \citep[e.g.,][]{shapley03}, and thus that is likely not the case here.  In addition, while the [Ne\,{\sc v}] emission is most likely due to an AGN, there is the possibility that it arises due to shocks, although none of our objects solely identified by this line inhabit the BPT plane (see \S~3.3) near the shock models of \citet{dopita95} \citep[c.f.,][]{vandokkum05}.

[Ne\,{\sc iii}] $\lambda$3869 is detected in numerous objects.  However, this line may not conclusively point to an AGN, as its ionization potential is not much greater than [O\,{\sc iii}].  As a control sample, we examined SDSS spectroscopic data\footnote[7]{http://www.mpa-garching.mpg.de/SDSS/}, to see where objects with detected [Ne\,{\sc iii}] emission inhabit the BPT diagram.  We found that at all fluxes, objects with [Ne\,{\sc iii}] emission populate both the star-forming and AGN sequences, implying that [Ne\,{\sc iii}] emission is not conclusive evidence of an AGN.

\subsection{Emission Line Diagnostics}
In Figure 2 we plot ratios of H$\alpha$/[N\,{\sc ii}] versus [O\,{\sc iii}]/H$\beta$ to diagnose the ionizing source within a galaxy \citep*{baldwin81}.  In this plane galaxies segregate nicely, with separate star-forming and AGN sequences.  The contours represent the two sequences as defined using $\sim$ 10$^{5}$ SDSS galaxies\footnote[8]{From the CMU-Pitt Value Added Catalog; http://nvogre.phyast.pitt.edu/vac/}, with the lighter contours representing the star-forming sequence, and the darker contours representing the AGN sequence.  We plot each of our objects\footnote[9]{EGS27 cannot be plotted in the plane as H$\alpha$ was shifted out of the observable range.} on this plane using the various colored symbols.  We also plot the theoretical maximum starburst line from \citet{kewley01} and the star formation/AGN demarcation line from \citet{kauffmann03c}.  We estimated flux limits for undetected lines by placing mock emission lines in the spectrum, and decreasing the line strength until the detection significance dropped below 3 $\sigma$, finding that 1 $\sigma$ line flux limits were typically $\approx$ 1.0 $\times$ 10$^{-17}$ erg s$^{-1}$ cm$^{-2}$.  In objects where any of these lines was undetected, we plot their 3 $\sigma$ upper-limits.
\begin{figure*}
\epsscale{1.0}
\plotone{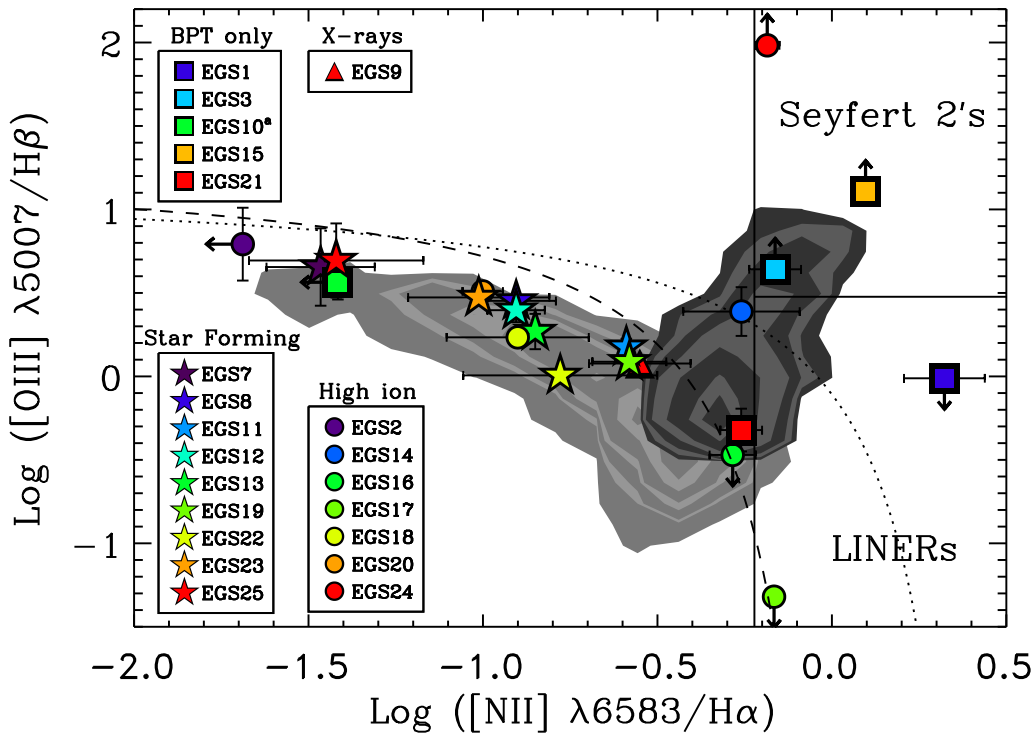}
\caption{Our 23 LAEs plotted on the standard BPT line ratio diagram.  3 $\sigma$ upper limits are plotted for undetected lines.  The light and dark gray contours represent the star-forming and AGN sequences, respectively, from SDSS data, with the star-forming contours representing 2, 15, 25, 60, 100, 300 and 500 galaxies per 0.03 $\times$ 0.03 dex bin, and the AGN contours representing 5, 8, 12, 20, 40, 75 and 100 AGNs per bin.  The dotted line is the maximum starburst curve from Kewley et al.\ (2001), while the dashed curve is an updated SF/AGN demarcation line from \citet{kauffmann03c}.  The solid lines denote regions expected to be inhabited by Seyfert 2's and LINERs \citep{kauffmann03c}. $^{a}$EGS10 was identified as a possible AGN from the BPT O\,{\sc i} $\lambda$6300/H$\alpha$ versus [O\,{\sc ii}]/[O\,{\sc iii}] plane.}\label{fig:bpt}
\end{figure*}

EGS1, EGS3, EGS14, EGS15 and EGS24 appear to unambiguously lie on the AGN sequence, and we henceforth classify them as AGN.  EGS24 has already been noted for its broad emission lines, as well as its [Ne\,{\sc v}] and He\,{\sc ii} emission, and EGS14 also has a 2.9 $\sigma$ [Ne\,{\sc v}] detection.  Although EGS1, EGS3 and EGS15 were not indicated by either of our previous classification methods, they are located very far from the star-forming sequence, and significantly higher than the maximum starburst curve.  However, EGS1 appears to be in the LINER regime, and thus if an AGN is present, it may not be dominating the observed continuum and line fluxes.  EGS16, EGS17 and EGS21 lie in regions which could correspond to either AGN or star-forming galaxies, as well as near the Kauffmann et al.\ demarcation line.  EGS16 and EGS17 have detected [Ne\,{\sc v}] emission.  As EGS21 was not indicated by any of our previous methods, thus we also only consider it a possible AGN.

\citet*{baldwin81} also derived classifications using lines of O\,{\sc i} and [O\,{\sc ii}], and their Fig. 4, which plots O\,{\sc i} $\lambda$6300/H$\alpha$ versus [O\,{\sc ii}]/[O\,{\sc iii}], shows a clear separation in O\,{\sc i} $\lambda$6300/H$\alpha$, with nearly all power-law ionization objects (i.e. AGN) having log(O\,{\sc i}/H$\alpha$) $\gtrsim$ -1, and all star-forming objects having log(O\,{\sc i}/H$\alpha$) $\lesssim$ -1.5.  We find that out of the five objects which have $\geq$ 3 $\sigma$ detections of O\,{\sc i} $\lambda$6300, EGS10 and EGS24 exhibit log(O\,{\sc i}/H$\alpha$) indicative of AGN activity, of -0.74 and -1.05, respectively.  EGS10 was not indicated by any of our previous classification methods, thus we now regard it as a possible AGN.

\subsection{Mid-Infrared}
\citet{stern05} defined a color-color region using colors from the {\it Spitzer} Infrared Array Camera (IRAC) which preferentially selects AGN in the redshift range of our sample (20\% contamination), due to the rising power law spectrum (i.e. from warm dust) exhibited by these objects.  We examined the AEGIS IRAC catalog and found that four of our objects were covered by the IRAC data: EGS1, EGS9, EGS15 and EGS27.  All four were detected in all IRAC bands, while only EGS27 lies within the selection wedge, at $\sim$ 1 $\sigma$ from the edge.  Thus, we regard EGS27 as a possible AGN.  While other methods select the remaining three objects as AGN, they all reside outside the wedge.
\begin{deluxetable}{ccccccc}
\tablecaption{Measured Object Properties}\label{tab:final}
\tablewidth{0pt}
\tablehead{
\colhead{Object} & \colhead{ID} & \colhead{Redshift} & \colhead{FWHM$_{P}$} & \colhead{24 $\mu$m Flux} & \colhead{AGN} & \colhead{Class}\\
\colhead{$ $} & \colhead{$ $} & \colhead{$ $} & \colhead{(km s$^{-1}$)} & \colhead{(mJy)} & \colhead{Grade} & \colhead{Method}\\
}
\startdata
EGS1\phantom{0}&32462&0.1996&---&---&B&3\\
EGS2\phantom{0}&7430\phantom{0}&0.2078&\phantom{0}221 $\pm$ \phantom{00}1&0.23 $\pm$ 0.02&B&2\\
EGS3\phantom{0}&5087\phantom{0}&0.2102&\phantom{0}155 $\pm$ \phantom{0}48&---&B&3\\
EGS7\phantom{0}&18322&0.2440&\phantom{0}220 $\pm$ \phantom{00}1&0.16 $\pm$ 0.01&SF&---\\
EGS8\phantom{0}&2682\phantom{0}&0.2395&\phantom{0}206 $\pm$ \phantom{00}5&---&SF&---\\
EGS9\phantom{0}&5715\phantom{0}&0.2462&\phantom{0}276 $\pm$ \phantom{00}6&0.20 $\pm$ 0.01&B&4\\
EGS10&17005&0.2466&\phantom{0}223 $\pm$ \phantom{00}3&---&C&3\\
EGS11&4719\phantom{0}&0.2524&\phantom{0}189 $\pm$ \phantom{00}4&0.14 $\pm$ 0.01&SF&---\\
EGS12&21404&0.2515&\phantom{0}194 $\pm$ \phantom{00}3&---&SF&---\\
EGS13&12279&0.2607&\phantom{0}234 $\pm$ \phantom{00}2&0.30 $\pm$ 0.03&SF&---\\
EGS14&14069&0.2578&\phantom{0}229 $\pm$ \phantom{0}13&---&A&2,3\\
EGS15&21024&0.2630&\phantom{0}174 $\pm$ \phantom{0}29&0.23 $\pm$ 0.02&B&3\\
EGS16&3488\phantom{0}&0.2643&\phantom{0}211 $\pm$ \phantom{0}12&---&A&2,3\\
EGS17&3525\phantom{0}&0.2647&\phantom{0}244 $\pm$ \phantom{0}13&0.62 $\pm$ 0.06&A&2,3\\
EGS18&29573&0.2689&\phantom{0}200 $\pm$ \phantom{0}10&---&B&2\\
EGS19&31403&0.2666&\phantom{0}200 $\pm$ \phantom{00}3&---&SF&---\\
EGS20&33559&0.2680&\phantom{0}210 $\pm$ \phantom{00}3&---&C&2\\
EGS21&9045\phantom{0}&0.2814&\phantom{0}229 $\pm$ \phantom{00}7&---&C&3\\
EGS22&21579&0.2828&\phantom{0}245 $\pm$ \phantom{00}9&0.16 $\pm$ 0.01&SF&---\\
EGS23&28751&0.2865&\phantom{0}264 $\pm$ \phantom{00}2&---&SF&---\\
EGS24&23096&0.3025&1064 $\pm$ \phantom{0}17&---&A&1,2,3\\
EGS25&29558&0.3243&\phantom{0}240 $\pm$ \phantom{00}1&---&SF&---\\
EGS27&10182&0.4512&\phantom{0}257 $\pm$ \phantom{00}2&0.29 $\pm$ 0.01&C&5\\
\enddata
\tablecomments{ID number is from Deharveng et al.\ (2008).  The redshift is the weighted mean of all lines detected at $\geq$ 3 $\sigma$ significance, with a characteristic error $\Delta$z $\sim$ .00001.  FWHM$_{P}$ represents the width of detected lines of hydrogen and helium.  The AGN classification grades refer to our level of confidence in the classification, and are detailed in \S 4.1.  The last column denotes the method of classification: 1=Line width, 2=Highly ionized line emission, 3=BPT line ratios, 4=X-rays and 5=IRAC color..
}
\end{deluxetable}

\subsection{X-rays}
The {\it Chandra} AEGIS-X 200 ksec catalog (v2.0; \citet{laird09}) has limiting soft and hard-band luminosities of L$_{0.5-2 keV (SB)}$ = 1.5 $\times$ 10$^{40}$ and L$_{2-10 keV (HB)}$ = 1.1 $\times$ 10$^{41}$ erg s$^{-1}$ (z = 0.3), respectively, thus if our objects had typical AGN X-ray luminosities, we should be able to detect them.  We examined the positions of each of the 11 objects which were covered by the {\it Chandra} observations, and found that only EGS9 has a X-ray counterpart within 5\arcs (a separation of r $<$ 0.19\arcs), with full-band luminosity L$_{FB}$ = 8.0 $\pm$ 1.5 $\times$ 10$^{41}$ erg s$^{-1}$ (at z = 0.246).  While star-forming galaxies typically show L$_{FB}$ $<$ 10$^{42}$ erg s$^{-1}$ \citep{szokoly04}, EGS9's hardness ratio of -0.13, which when combined with the X-ray luminosity, satisfies the \citet{szokoly04} AGN criteria, thus we regard EGS9 as a probable AGN.  The hardness ratio is consistent with the emission detected at 24 $\mu$m (see \S 3.5), which could imply that dust is obscuring some of the X-rays.  An object similar to EGS9 at high-redshift likely would not be selected as an AGN via its X-ray flux, as its X-ray/Ly$\alpha$ flux ratio of $\sim$ 1.2 is much less than the typical limit from the Large Area Lyman Alpha (LALA; \citet{rhoads00}) survey of $>$ 10 \citep{wang04}.

Out of the 10 X-ray undetected LAEs, our previous methods have found possible AGNs in six of them.  This can be understood given the unification picture of AGN \citep[e.g.,][]{urry95}, where one has a direct line-of-sight to see the X-rays from the accretion disk.  When observing from other angles, the accretion disk can be blocked by the torus of dust and gas surrounding the center.  In this case, the dust emission from the torus may be re-emitted at long wavelengths, thus we examined the MIPS 24 $\mu$m catalog from the Far-Infrared Deep Extragalactic Legacy survey (Dickinson et al.\ 2009, in prep).  Out of these six LAEs, we found that three had a MIPS counterpart (EGS2, EGS15 and EGS27; all 9 MIPS detections out of the 13 LAEs covered are listed in Table 1), indicating that dust may be absorbing the X-rays.

To see whether any of the six X-ray undetected LAEs classified as AGNs harbor low-luminosity AGNs, we performed an X-ray stacking analysis.  We found that the stack was marginally detected with a S/N $\sim$ 2.1.  Assuming $\Gamma$ = 1.4, the mean X-ray luminosity of the individually detected objects is L$_{SB}$ (L$_{HB}$) = 1.39 $\pm$ 0.67 (1.28 $\pm$ 2.82) $\times$ 10$^{40}$ erg s$^{-1}$ at the median redshift of these six objects (z = 0.263).  The S/N decreases to $\sim$ 1.8 when stacking all 10 X-ray undetected LAEs.  This shows that the six AGNs are dominating the X-ray flux from all 10 undetected objects, and that they likely harbor low-luminosity AGNs, which are too faint to be individually detected in the AEGIS-X survey.  Our NLAGN number density (see \S 4.1) of 7 $\times$ 10$^{-5}$ Mpc$^{-3}$ is consistent with this conclusion, as \citet{hasinger05} show that the number density of AGNs increases with decreasing luminosity, and n $\sim$ 1 $\times$ 10$^{-5}$ Mpc$^{-3}$ in their lowest luminosity bin (10$^{42}$ $\leq$ L$_{x}$ $\leq$ 10$^{43}$ erg s$^{-1}$).

\section{Discussion}
\subsection{Final Classification}
We assign confidence levels to the classified AGNs, with objects indicated by two or more methods given a classification grade of ``A''; objects with unambiguous classification via only one method are given a grade of ``B''; and objects with a possible classification with one method are given a grade of ``C.''  As shown in Table~1, we have 4 grade A, 6 grade B and 4 grade C AGNs.  Taking A \& B classifications as being reliably indicative of AGN activity, we find an AGN fraction of 43\% (10/23) in low-redshift LAEs.  Including class C AGNs, this could be as high as 61\% (14/23), while neglecting class B \& C AGNs provides a minimum value of 17\% (4/23).  A classification of A or B calls into doubt the stellar population modeling results of \citet{finkelstein09c} for that particular object.

Applying our AGN fraction (0.435) and SF fraction (0.565) to the 38/39 EGS LAEs at 0.2 $\leq$ z $\leq$ 0.35 from \citet{deharveng08}, we find total numbers of: 38 total LAEs, 16.5 LAEs with NLAGN and 21.5 SF-dominated LAEs, corresponding to number densities of: 1.6 $\times$ 10$^{-4}$ Mpc$^{-3}$, 7.0 $\times$ 10$^{-5}$ Mpc$^{-3}$ and 9.1 $\times$ 10$^{-5}$ Mpc$^{-3}$, respectively.  From their photometric sample, \citet{ouchi08} found LAE number densities of 5.1, 1.6, and 4.3 $\times$ 10$^{-4}$ Mpc$^{-3}$ at z = 3.1, 3.7 and 5.7, respectively.  Comparing to our number density for all star-forming LAEs, we find that on average LAEs are $\sim$ 4 times rarer than at high redshift.  This is a rough comparison, as \citet{deharveng08} is a factor of $\sim$ 7 more sensitive in luminosity to typical high-redshift studies.  On the flip side, many of the AGNs here have low L$_{x}$, and thus would not be flagged as AGNs at high redshift.

\subsection{Evolution in LAE AGN Fraction?}
Using X-ray data, typical AGN fractions in LAE samples at z $>$ 3 range from $\sim$ 0 -- 1\% \citep{malhotra03,wang04,gawiser07,ouchi08}, increasing to 5\% at z $\sim$ 2.25 \citep{nilsson09}, implying that the AGN fraction in LAEs may be evolving below z = 3.  However, these studies only probed intermediate X-ray luminosities, thus low-luminosity AGN like we find in this Letter may have been missed.  \citet{dawson04} and Wang et al.\ (2009) examined z $\sim$ 4.5 LAEs with rest-frame ultraviolet spectroscopy, and also found little evidence for AGN activity, with Wang et al.\ reporting an upper limit on the AGN fraction of $<$ 17\% from the upper limit on the C\,{\sc iv} $\lambda$1549 emission line.  Line widths can also be used at high redshift, however BLAGN appear to be rare in high-redshift LAEs \citep[e.g.,][]{dawson04}.

Our AGN fraction of 43 $^{+18}_{-26}$\% for z $\sim$ 0.3 LAEs is significantly higher than high-redshift measurements.  While at first glance this is a fair comparison, as high-redshift LAEs lack BLAGNs, and \citet{deharveng08} removed all BLAGNs from their sample (8 out of 47 original EGS LAEs were removed as BLAGNs, corresponding to n$_{BLAGN}$ = 2.1 $\times$ 10$^{-5}$ Mpc$^{-3}$ over 0.2 $\leq$ z $\leq$ 0.35; J.-M Deharveng, private communication), only three of our methods (line widths, X-rays and high-ionization emission) have been used in LAEs at high redshift.  With only these methods we obtain an AGN fraction of 26\%, better suited to comparisons with high-redshift.  When restricted to measurements available at high-redshift, we thus miss from $\sim$ 15 - 40\% of the AGNs in our sample, implying that high-redshift LAE AGN fraction measurements may be lower limits.

The AGN X-ray luminosities in our sample are $\sim$ 100 times lower than current sensitivities at high redshift.  Thus it is unclear if the LAE AGN fraction is evolving: If the low-luminosity AGN we detect at z $\sim$ 0.3 also exist at high redshift, we cannot yet detect them (although the increase in LAE AGN fraction at z $\sim$ 2 compared to z $>$ 3 may be due to the decreased luminosity distance).  Given the observed downsizing trend of AGN luminosity with decreasing redshift \citep[e.g.,][]{cowie03}, the higher LAE AGN fraction at z $\sim$ 0.3 is likely a combination of evolution, coupled with the presence of undetected low-luminosity AGN at high redshift.  Even if a substantial number of low-luminosity AGNs similar to their z $\sim$ 0.3 cousins exist at high redshift, their contribution to the total \lya~flux is minor, scaling from the X-ray luminosity and the \lya~to X-ray ratios of Seyferts \citep{kriss84}.  This implies the \lya~flux is star-formation dominated.  Nonetheless, infrared spectroscopy and deeper X-ray imaging is necessary to detect the true high-redshift LAE AGN fraction, and to quantify the AGN contamination.  Some of this capability will be available in the not too distant future with the {\it James Webb Space Telescope}.

\acknowledgements
We thank the referee for their suggestions which improved the clarity of the discussion.  We also thank Jean-Michel Deharveng and Eric Gawiser for helpful discussions.  This work was supported by GALEX Archival Grant \#GI5-036 and the Texas A\&M University College of Science and Department of Physics.

\end{document}